\documentclass[12pt,a4paper]{article}

\usepackage[english]{babel}
\usepackage{float}
\usepackage{booktabs}
\usepackage{amsmath}
\usepackage{rotating}
\usepackage{graphics}
\usepackage{listings}
\usepackage{setspace}
\usepackage{natbib}
\bibpunct{(}{)}{;}{a}{,}{,}
\usepackage[titletoc]{appendix}
\usepackage{cite}
\usepackage[left=1.5cm,right=1.5cm,top=2cm,bottom=2cm]{geometry}
\usepackage[titletoc]{appendix}
\usepackage[font=footnotesize, labelfont={bf}, margin=1cm]{caption}
\usepackage{rotating}
\usepackage{tabularx}
\usepackage{array}
\usepackage{colortbl}
\usepackage{enumerate,multicol}
\usepackage{amsmath, amssymb, amsthm}
\usepackage{enumitem,bm}
\usepackage{slashbox}
\newcolumntype{L}[1]{>{\raggedright\let\newline\\\arraybackslash\hspace{0pt}}m{#1}}
\newcolumntype{C}[1]{>{\centering\let\newline\\\arraybackslash\hspace{0pt}}m{#1}}
\newcolumntype{R}[1]{>{\raggedleft\let\newline\\\arraybackslash\hspace{0pt}}m{#1}}
\usepackage[para,online,flushleft]{threeparttable}
\usepackage{parcolumns}

\usepackage{standalone}
\providecommand{\keywords}[1]{{\bf{Key words:}} #1}
\providecommand{\tables}[1]{{\bf{Tables:}} #1}
\providecommand{\figures}[1]{{\bf{Figures:}} #1}
\providecommand{\words}[1]{{\bf{Words:}} #1}
\providecommand{\runhead}[1]{{\bf{Running Head:}} #1}
\usepackage[multiple]{footmisc}
\usepackage[sectionbib]{chapterbib}
\usepackage{etoolbox}
\usepackage{authblk}
\usepackage[position=bottom]{subfig}
\usepackage{pdfpages}

\begin{document}

\title{Handling Missing Data in Within-Trial Cost-Effectiveness Analysis: a Review with Future Recommendations}
\author[1]{Andrea Gabrio\thanks{E-mail: andrea.gabrio.15@ucl.ac.uk}}
\author[2]{Alexina Mason}
\author[1]{Gianluca Baio}
\affil[1]{\small \textit{Department of Statistical Science, University College London, 1-19 Torrington Place, London WC1E 7HB, UK.}}
\affil[2]{\small \textit{Department of Health Services Research and Policy, London School of Hygiene and Tropical Medicine, 15-17 Tavistock Place, London WC1H 9SH, UK.}}

\date{}
\maketitle

\hrule
\abstract{Cost-Effectiveness Analyses (CEAs) alongside randomised controlled trials (RCTs) are increasingly often designed to collect resource use and preference-based health status data for the purpose of healthcare technology assessment. However, because of the way these measures are collected, they are prone to missing data, which can ultimately affect the decision of whether an intervention is good value for money. We examine how missing cost and effect outcome data are handled in RCT-based CEAs, complementing a previous review (covering 2003-2009, 88 articles) with a new systematic review (2009-2015, 81 articles) focussing on two different perspectives. First, we review the description of the missing data, the statistical methods used to deal with them, and the quality of the judgement underpinning the choice of these methods. Second, we provide guidelines on how the information about missingness and related methods should be presented to improve the reporting and handling of missing data. Our review shows that missing data in within-RCT CEAs are still often inadequately handled and the overall level of information provided to support the chosen methods is rarely satisfactory.}
\newline\newline
\newline\newline
\tables{1} \figures{5} \words{4900}
\newline
\runhead{Handling missing data: a review and guidelines}

\newpage

\section{Introduction}
A well-known issue in Cost-Effectiveness Analysis (CEA), especially within a Randomised Controlled Trial (RCT) setting, is the presence of large proportions of missing data in either or both the outcome variables, i.e.~the cost and the clinical effectiveness or utility measures. Removing the unobserved cases (a method usually referred to as ``Complete Case Analysis'', CCA) generally leads to a loss in efficiency and possible serious biases in the parameter estimates \citep{Rubina, Schafera, Littlea, Molenberghs}. Consequently, the final conclusions of the study may be strongly influenced by the way in which missingness is handled, thus possibly reversing the decision about the cost-effectiveness of a new treatment compared to the standard option \citep{Manca, Marshall}.

While the problem of missing data is widely discussed in the statistical literature, it has been relatively overlooked in the health economics one. Notable exceptions include \citet{Graves,Briggs,Oostenbrink, Burton, Lambert}, mainly focussing on the cost measures; \citet{Richardson,Wood, Groenwold, Powney,Simons, Rombach}, with reference to health outcome measures; and \citet{Manca, Harkanen, Diaz-Ordaz, Faria}, who consider both outcomes. 

Interestingly, recent reviews on the methods applied in within-trial CEAs \citep{Noble, Diaz-Ordaz2} have concluded that CCA has historically represented the standard approach in health economics. As a result, we should be naturally sceptical about the conclusions achieved by CEAs performed in a context where missingness is not addressed in a principled way. This implies the incorporation of uncertainty about the missing values by combining available information from the observed data with statistical assumptions to build a well-defined statistical model. Within this framework, subsequent inferences are valid under these assumptions, which in turn can be varied to test their impact on the decision-making.

The objective of this article is twofold: first, we review the methods used to handle missingness in within-trial CEAs between 2003-2015 by updating and extending the work of \citet{Noble}. This is done with a view to assessing whether the methods have evolved over time. Second, we provide some guidelines about the way in which missingness should be analysed and reported in the studies. The paper is structured as follows: \S\ref{mechanisms} illustrates Rubin's classification of missing data mechanisms \citep{Rubina}. In \S\ref{handling} we provide a brief summary of the most popular missingness methods, while \S\ref{review} presents the methodology used to select the review's articles and the main results derived from the different analyses performed. In \S\ref{analysis} we report a qualitative analysis of the way in which information about the missing data is provided in the studies and use these results to form some guidelines on the approach authors should follow when dealing with missingness in CEAs. Finally, \S\ref{conclusions} summarises our findings and recommendations.

\section{Missing Data Mechanisms}\label{mechanisms}
When analysing partially observed data, it is essential to investigate the possible reasons behind missingness. This formally translates into an \textit{assumed} missing data mechanism that is linked to the data generating process.

We consider a sample of $i=1,\ldots,n$ individuals and for each the relevant outcome is indicated as $y_i$, which is unobserved for some individuals. Typically, trial data also include a set of covariates $\bm{x}_i=(x_{1i},\ldots,x_{Ji})$, e.g.~sex, age or co-morbidities. While in general these may be partially or fully observed, in this section we consider only the latter case. In addition, we define a missingness indicator $m_{i}$ taking value 1 if the $i-$th subject is associated with missing outcome and 0 otherwise. 

This setting can be modelled using two sub-models, or ``modules''. The first module is the missing data mechanism, denoted as \textit{Model of Missingness} (MoM). It describes a probability distribution for $m_{i}$, as a function of some unobserved parameters $\pi_{i}$ and $\delta$, defining the probability of missingness in the outcome variable $y_{i}$. The second module is the data generating process of the outcome variable, denoted as \textit{Model of Analysis} (MoA). This contains the main parameters of interest (e.g.~the population average costs and benefits) and describes a probability model for the outcome $y_{i}$. As a general example, we can think of a simple regression model where $y_i\sim\mbox{Normal}(\mu_i,\sigma)$, and $\mu_{i} = \beta_0+\beta_1 x_i$. In this case, the parameters of the MoA are $\bm\beta=(\beta_0,\beta_1)$ and $\sigma$.

The most accepted classification of missing mechanisms is given by \citet{Rubina} and is based on three classes, according to how the missingness probability in the MoM is modelled. A simple graphical representation of the three classes is provided in Figure \ref{FMDM}. Variables and parameters are denoted by nodes of different shapes and colours according to their nature. Parameters ($\beta_0$, $\beta_1$, $\sigma$, $\delta$) are represented through grey circles. ``Logical'' quantities such as $\mu_{i}$ and $\pi_{i}$, which are defined as function of the parameters, are denoted by a double circle notation. Fully observed variables ($m_{i}$) are represented with a white circle while partially observed variables ($y_{i}$) are denoted by a darker grey circle.  Finally, we show covariates ($x_{i}$) as white squares to indicate that they are fully observed and not modelled. Rounded rectangles are used to show the extent of the two modules in terms of variables/parameters included. Arrows show the relationships between the nodes, with dashed and solid lines indicating logical functions and stochastic dependence, respectively.

\begin{figure*}[!ht]
\centering
FIGURE 1 HERE
\end{figure*}

Figure \ref{FMDM} (a) illustrates  the class of `Missing Completely At Random' (MCAR), in which the probability of missingness is fully independent of any other partially or fully observed variable. Consequently, in Figure \ref{FMDM} (a) MoA and MoM are not connected and $\pi_{i}$ does not depend on any quantity in the MoA. This amounts to assuming that there is no systematic difference between partially and fully observed individuals in terms of the outcome $y_{i}$. In other words, in this case we would be assuming that observed cases are a representative sample of the full sample.

Figure \ref{FMDM} (b) shows a case of `Missing At Random' (MAR), in which the missingness probability may depend on a fully observed variable. As a result, MoA and MoM are connected by means of the predictor variable affecting both the mechanisms generating $y_{i}$ and $m_{i}$. Because of this relationship, the partially observed cases are systematically different from the fully observed cases; crucially, however, the difference is fully captured~by~$x_{i}$.  

Figure \ref{FMDM} (c) provides an example of `Missing Not At Random' (MNAR). This is characterised by dependence of the probability of missingness on both the partially and fully observed variables. Thus, in Figure \ref{FMDM} (c) $\pi_{i}$ depends on both the fully observed predictor $x_{i}$ and the partially observed outcome $y_{i}$. This means that the difference between fully and partially observed cases still depends on the missing values, even after taking $x_{i}$ into account. Therefore it is necessary to make more structured assumptions about this relationship that go beyond the information contained in the data. 

While intuitively helpful, this framework may be too simplistic in some cases. Since the scope of this section is to provide a broad overview for Rubin's classification, we assumed the simplest case where missingness is present in a single response variable only, which may not hold in real applications. This is particularly likely in the context of CEA, in which we are concerned with a multivariate outcome, made of suitable measures of clinical benefits and costs, i.e.~$y_i=(e_i,c_i)$. Missingness can occur for either or both the relevant outcomes and this can lead to as many missingness mechanisms as the number of partially observed quantities (covariate missingness must also be considered). Additional complexity is given by whether data are obtained in a cross-sectional or longitudinal setting, static or time-varying covariates and more importantly the possible correlation between variables and missingness mechanisms and between the mechanisms themselves. 



\section{Methods to Handle Missing Data}\label{handling}
There are many different statistical methods to account for missingness, each relying on different assumptions. It is important to carefully select the method in line with the setting-specific assumptions we assume to hold. For the sake of simplicity here we only broadly categorise these methods. More in depth and complete presentation and analysis can be found for example in \citet{SchaferMD, Molenberghs}.

\subsection{Complete Case Analysis}
This is a popular method in within-trial CEA studies, despite its limitations due to the strong assumption that only the fully observed cases are needed in order to correctly make inference. The critical disadvantage is that missing cases are simply discarded, thus reducing efficiency and possibly biasing the parameter estimates.

\subsection{Single Imputation}
Single Imputation (SI) methods replace the missing data with a single predicted value, such as the unconditional or conditional mean or the last value observed for a given case. This category includes Last Value Carried Forward \citep{Shao}, Linear Extrapolation \citep{Twisk} and Conditional Imputation \citep{Buck}. Although sometimes valid, these methods are never recommended as they typically require stronger assumptions than MCAR and always fail to take account of the uncertainty underlying the imputation process, i.e.~they do not recognise that the imputed values are \textit{estimated} rather than \textit{known}.

\subsection{Multiple Imputation}
A more sophisticated method is Multiple Imputation \citep[MI,][]{SchaferMI}. The underlying idea is to fill-in each missing data with plausible simulated values, drawn from the conditional predictive distribution of the missing given the observed values. Thus, the set of imputations can properly represent the information about the missing values that is contained in the observed data for the chosen model. This is repeated $K$ times, leading to $K$ imputed datasets that can be analysed via complete-data methods. The individual estimates are then combined into a single quantity, e.g.~using Rubin's rules \citep{Rubina}; this captures the variability within and between imputations. However, the critical aspect is that valid inferences depend on the correct specification of the imputation model in terms of variable selection, distributions and correlations.

\subsection{Sensitivity Analysis}\label{sa}
Sensitivity Analysis (SA) is a technique used to determine how different input values in a model will impact the output, under a given set of assumptions. When applied to missing data this corresponds to exploring as many plausible missing data assumptions as possible and then assessing how consistent results are across the different scenarios. In particular, it is generally recommended to set MAR as the reference assumption and then explore different MNAR departures from MAR, to assess the robustness of the results to different plausible alternative missingness mechanisms. The purpose of such analysis is to account more fully for the uncertainty about the missingness. Usually SA is implemented through more advanced methods that are able to explicitly model a MNAR mechanism such as Selection or Pattern Mixture Models \citep{Molenberghs,Daniels}.

\subsection{Others}
There is a wide list of possible alternatives that could be explored such as Inverse Probability Weighting \citep{Robins2}, Likelihood-based methods \citep{McLachlan}, Doubly Robust \citep{Bang} or Full Bayesian methods \citep{Daniels,Mason}.

\section{Literature Review}\label{review}

\subsection{Methods}
\citet{Noble} (henceforth NHT) reviewed the methods used to handle missing cost measures in 88 articles published during the period 2003-2009. We extend their review, to include missing effects. Further, we use NHT's strategy to identify papers in the subsequent period, 1 April 2009 to 31 December 2015. Articles were considered eligible for the review only if they were cost-effectiveness analyses within RCTs, used individual patient-level data and mentioned \texttt{missing data} in the text. We relied on the search engines of three online full-text journal repositories: \texttt{Science-Direct.com}, \texttt{bmj.com}, and The Database of Abstracts of Reviews of Effects (DARE) and NHS Economic Evaluation Database (NHS EED). The key words used in the search strategy were (\texttt{cost effectiveness} OR \texttt{economic evaluation}) AND \texttt{missing data} AND \texttt{trial} AND (\texttt{randomised} OR \texttt{randomized}). The on-line databases identified 1129 articles most of which were duplicates. After abstract review, 128 articles were considered, of which 81 fulfilled the eligibility criteria. 

We present and compare the articles reviewed for the two periods by type of analysis performed. First, we look at the base-case methods implemented, i.e.~those used in the main analysis embedding the assumptions about missing data. Second, we consider any alternative methods discussed; when present, these assess the robustness of the results obtained in the main analysis against departures from the initial assumptions on missingness. 

\subsection{Base-case Analysis}
As shown in Figure \ref{F2} (a), NHT found that CCA was the most popular base-case method, used in $31\%$ of the papers; 23\% were unclear about the technique adopted. Single imputation methods were well represented, with mean imputation and conditional imputation used in $10\%$ and $9\%$ of the articles respectively. MI was found in $9\%$ of the articles. Our analysis of the methods for missing effectiveness measures shows a similar pattern in Figure \ref{F2} (c). CCA was used in $27\%$ of the cases and with a sizeable proportion of papers unclear about the technique adopted ($24\%$). Single imputation methods are here dominated by LVCF ($10\%$), while a slightly higher proportion uses MI ($15\%$).

In 2009-2015, MI replaces CCA as the most frequently used base-case method in both costs and effects, at $33\%$ and $34\%$ respectively (Figures 2 (b) and (d)). However, CCA is still the method of choice in many papers ($15\%$ for costs and $21\%$ for effects). The proportion of papers that are unclear about the chosen method is similar over the two time periods for costs, but halves in the later period for effects.

\begin{figure*}[!ht]
  \centering
  FIGURE 2 HERE
\end{figure*}

\subsection{Robustness Analysis}

With the term robustness analysis we refer to the specific concept introduced in \S\ref{sa}, whose aim is to assess the impact of plausible alternative assumptions about the missing data on the results, with respect to the base-case scenario. This implies varying the structural assumptions about the missingness mechanisms underlying the model, as opposed to the more general definition of sensitivity analysis, which concerns with varying lower level assumptions (e.g.~about the distribution associated with a variable in the model).

Despite having a key role for assessing uncertainty, in practice, robustness analyses are rarely performed in CEAs. This poses an important question related to the reliability of the findings, as they may be affected by the specific assumption about missing data. From both review periods it seems that a robustness analysis is infrequently used and typically involves only one alternative scenario. This is not likely to be an optimal choice as the main objective of this analysis is to explore as many plausible alternative missing data assumptions as possible.

NHT found that $75\%$ (66/88) of the articles did not include any robustness analysis, with the remaining papers typically performing an analysis by comparing CCA and MI. Similar findings apply to missing effects, with about $76\%$ (67/88) of the studies lacking any alternative missing data method. Similarly in the 2009-2015 review, we observe no robustness analysis in the majority of the articles for both costs ($75\%$ or 61/81) and effects ($70\%$, 51/81).

Figure \ref{F3} provides a pictorial overview of the alternative methods used for cost and effect data. For costs most articles describe no alternative analysis. In the earlier period, the choice of alternative missingness methods seems well-spread across CCA, MI and the use of more than one method, with a slightly more frequent adoption of MI. By contrast, in the later period, more cases use CCA as a robustness method in combination with MI as the base-case method.

Figures \ref{F3} (c)--(d) describe the effects, with most of the articles not reporting any robustness analysis and with a significant increase in MI analyses, opposed to a decrease in CCA, between the two periods. We also observe a reduction over time in the number of unclear missing effects method analyses. Excluding this category, there is a similar pattern to the cost graphs towards CCA used as a robustness method in combination with MI as the base-case~method.


\begin{figure*}[!ht]
  \centering
  FIGURE 3 HERE
\end{figure*}

\section{Recommendations for Missing Data Analyses}\label{analysis}
We argue that in order to judge whether missing data in a CEA have been adequately handled, a full description of the missingness problem, details of the methods used to address it and a discussion on the uncertainty in the conclusions resulting from the missingness are required. With this in mind, we have assembled guidelines on how information relating to the missing data should be reported (Table \ref{T1}). We define three broad categories (Description, Method, Limitations). For each, information that we consider vital for transparency is listed under `key considerations', while other details that could usefully be provided as supplementary material are suggested under `optimal considerations'. Comparing the information provided in the articles in our review against this list, allows us to qualitatively assess the quality of the reporting of how missingness has been handled in CEA in the two time periods.

To gain a fuller understanding of the current state of play, we also classify the articles from the perspective of the strength of the assumptions about the missingness mechanism. This is related to the choice of method, since each is underpinned by some specific missing data assumption. We can view the quality judgement and strength of assumptions as two dimensions providing a general mapping of how the missingness problem is handled. This applies to both the level of knowledge about the implications of a given missingness assumption on the results and how these are translated into the chosen method. Details of our evaluation of both aspects are provided next, starting with the strength of assumptions.

\subsection{Quality Evaluation Scheme}
We group the methods into five categories, ordered according to the strength of the associated missingness assumptions. These are: \textit{Single Imputation} (SI), typically requiring stronger assumptions than MCAR to hold; \textit{Complete Case Analysis}, usually associated with MCAR; \textit{Multiple Imputation}, generally based on MAR; and \textit{Unknown} (UNK), a residual group in which we classify studies that do not explicitly mention the method used. We associate this class with the strongest level of assumptions, since the lack of any method description may implicitly suggest (over)confidence in a small effect of missingness on the results. By contrast, we define \textit{Sensitivity Analysis} (SA) as the least restrictive approach, which can assess the robustness of the results to different alternative missingness mechanisms.

\begin{figure*}[!ht]
  \centering
  FIGURE 4 HERE
\end{figure*}

Using the list of key considerations in Table~\ref{T1}, first we determined whether no (all key considerations absent), partial (one or more key considerations absent) or full (all key considerations present) information has been provided for each component. From this we computed a numerical score to summarise the overall information provided on missingness, weighting the components in a ratio of 3:2:1. Finally, we converted the scores into grades A-E. Figure \ref{F6} shows the process and weights used. Although the importance between the different components is subjective, we believe that the chosen structure represents a reasonable and relatively straightforward assessment scheme.

\begin{figure*}[!ht]
  \centering
  TABLE 1 HERE
\end{figure*}

The resulting scores can be interpreted qualitatively as follows: 
\begin{enumerate}[leftmargin=1.8cm,rightmargin=.5cm]
  \item[\textbf{A (12)\;\;\;}] The highest quality judgement, identified by the upper thicker blue path in Figure~\ref{F6}, including only those studies that simultaneously provide all the key considerations for all the components. It is the benchmark for a comprehensive explanation and justification of the adopted missing data method.
  \item[\textbf{B (9-11)}] Includes studies providing full details for either the description or the method and at least partial information for the other components. Studies with no information about the limitations are only included in this category if full detail is provided for both the other components. 
  \item[\textbf{C (6-8)\;\;}] Studies for which information about missingness is not well-spread across the components. All key considerations are provided either for the description or the method, but with only a partial or no content in the other components. 
  \item[\textbf{D (3-5)\;\;}] Indicates a greater lack of  relevant information about missingness. Despite possibly including key considerations on any of the components, the information provided will at most be partial for the description in which case it will be combined with a total lack of content on either the method or the limitations.  
  \item[\textbf{E (0-2)\;\;}] The worst scenario where the overall information about the missing data is considered to be totally unsatisfactory. No description is given and we can observe at most only some of the key considerations for the method.
\end{enumerate}

\subsection{Grading the articles}

Figure \ref{F5} gives a graphical representation of both aspects for the articles reviewed between 2009-2015 in terms of the assumptions and justifications (quality scores) on missingness. In both graphs, more studies lie in the lower than in the upper part, indicating that fewer studies can be classified as high quality in terms of the considerations about missingness. This is highlighted by a greater concentration of points at the bottom of the figures (grade E). As we move along the vertical axis, this tends to reduce up to the top level (grade A), where only very few cases are shown. Of particular interest is the (almost) total absence of articles that performed a sensitivity analysis (SA), clearly indicating very slow uptake of this technique.

A shift along the vertical axis in the graphs indicates an increase in the level of understanding about the implications on the results for different choices of the missing data assumptions. Therefore, we can argue that an upward movement in the plot will always improve the justification of a specific assumption. However, to be able to follow this path we may have to rely on more sophisticated methods that can match the given missingness assumption, i.e.~if we think our data are MNAR, then CCA assumptions are less likely to hold. The aim of an optimal analysis should be to select a method that can be fully justified by matching the description of the missing data problem to the assumptions underpinning the chosen method, i.e.~map onto the upper section of the graphs.

\begin{figure*}[!ht]
  \centering
  FIGURE 5 HERE
\end{figure*}


\section{Discussion}\label{conclusions}

The objective of this paper is to critically appraise the issue of missing data analysis in within-trial CEAs. In addition, we aim at providing a set of recommendations to guide future studies towards a more principled handling and reporting of missingness. It is important that assumptions about missing data are clearly stated and justified. A robustness analysis is also important, in order to explore the impact of plausible alternative missingness assumptions on the results of the CEA. Often, a variety of techniques and analyses are used but not reported because of space limits; on-line appendices and supplementary material could be used to report these alternatives.

\subsection{Review}  
Figure \ref{F2} highlights a shift in the most popular base-case missingness method from CCA to MI, between the two periods of the review. The reasons behind this change may be related to some drawbacks of CCA and the relatively recent wide development of software to perform MI.

First, even under a strong missing data assumption such as MCAR, CCA remains inefficient because it ignores the predictive information contained in the partially observed cases. Non-negligible rates of missingness on a few variables of interest  may cause large portions of the sample to be discarded. Second, CCA may cause serious biases in the parameter estimates. Indeed, the condition for validity of CCA does not fit neatly into Rubin's classes \citep{White} in the important cases when: missing data affect the covariates; or the partially observed outcome has a longitudinal nature. 


Arguably, a very important factor in the increasing popularity of MI is the recent availability of specific computer routines or packages \citep[e.g.\ STATA or R; see][]{VanBuuren}. This probably led to some abuse of the method as noted by \citet{Molenberghs}. On the one hand, MI generally allows the inclusion of a larger number of variables/predictors in the imputation model than used in the analysis model, which potentially makes the assumption of MAR more plausible and thus the overall analysis less likely to be biased. On the other, the performance of MI depends on the correct specification of the imputation model (i.e.\ complexity in the analysis model is reflected in the imputation model) and care is required in its construction. Although essential, these details can be overlooked and are not often included in the reporting of the analysis, undermining its reliability.

From the comparison of the base-case methods used for the costs and effects between 2009 and 2015 (Figure \ref{F2}), we observe a marked reduction in the number of methods not clearly described for the effects, compared to those for the costs. In CEAs we typically observe effect data to be characterised by higher missing proportions than cost data, probably due to the different methodology used to collect them. While clinical effectiveness measures are usually collected through self-reported questionnaires, which are naturally prone to missingness, cost measures rely more on clinical patient files which may ensure a higher completeness rate. In addition, clinical outcomes are almost invariably the main objective of RCTs and as such they are usually subject to more advanced and standardised analyses. Arguably, costs are often considered as an add-on to the standard trial: for instance, sample size calculations are almost always performed with the effectiveness measure as the only outcome of interest. Consequently, missing data methods are less frequently well thought through for the analysis of the costs.

Our review identified only a few articles using more than one alternative method (Figure \ref{F3}). This situation indicates a gap in the literature associated with an under-implementation of robustness analyses, which may significantly affect the whole decision-making process outcome, under the perspective of a body who is responsible for providing recommendations about the implementation of alternative interventions for health care matters. Limiting the assessment of missingness assumptions to a single case is unlikely to provide a reliable picture of the underlying mechanism. This, in turn, may have a significant impact on the CEA and mislead its conclusions, suggesting the implementation of non-cost-effective treatments. Robustness analysis represents an important tool to properly account for more structured uncertainty related to the missing data and its implementation may provide a more realistic picture of the impact that the assumptions have on the final conclusions.


\subsection{Guidelines}
Generally speaking, most papers in our review achieved an unsatisfactory quality score under our classification (Figure \ref{F5}). Indeed, our benchmark area on the top-right corner of the graphs, is barely reached by less than $7\%$ of the articles, both for cost and effect data. The opportunity of reaching such a target might be precluded by the choice of the method adopted, which may not be able to support less restrictive assumptions about the missingness, even when this would be desirable. As a result, when simple methods cannot be fully justified it is necessary to replace them with more flexible ones that can relax assumptions and incorporate more alternatives. In settings such as those involving MNAR, sensitivity analysis might represent the only possible approach to account for the uncertainty due to the missingness in a principled way. However, due to the lack of studies either performing a SA or providing high quality scores on the assumptions, we argue that missingness is not adequately addressed in most studies. This could have the serious consequence of imposing too restrictive assumptions about missingness and affect the outcome of the decision making process.

The classification of the studies into ordered categories (Figure \ref{F6}) according to the information provided on missing data (Table \ref{T1}) is potentially a valuable tool for meta-analysis. It may be reasonable for analysts to assign different weights to the individual studies based on their specific information provided and methods adopted.

\subsection{Conclusions}   
Given the common high proportion of missing cost and effect data in within-trial CEAs, many study conclusions could be based on imprecise economic evidences. This is a potentially serious issue for bodies such as the National Institute for Health and Care Excellence (NICE) who use these evaluations in their decision making, thus possibly leading to incorrect policy decisions about the cost-effectiveness of new treatment options.

Our review shows, over time, a significant change from more to less restrictive methods in terms of the assumptions on the missingness mechanism. This is an encouraging movement towards a more suitable and careful missing data analysis. Nevertheless, improvements are still needed as only a small number of articles provide transparent information or perform a robustness analysis.

Our guidelines could represent a valuable tool to improve missing data handling. By carefully thinking about each component in the analysis we are forced to explicitly consider all the assumptions we make about missingness and assess the impact of their variation on final conclusions. The main advantage is a more comparable formalisation of the uncertainty as well as a better indication of possible issues in assessing the cost-effectiveness of new treatments.

\subsection*{Acknowledgements}
Dr Gianluca Baio is partially funded by a research grant sponsored by Mapi.\\
PhD student Andrea Gabrio is partially funded by a research grant sponsored by The Foundation BLANCEFLOR Boncompagni Ludovisi, n\'{e}e Bildt.\\


\clearpage

\begin{figure}[!h]
\centering
\subfloat[Missing Completely At Random (MCAR)]{\includegraphics{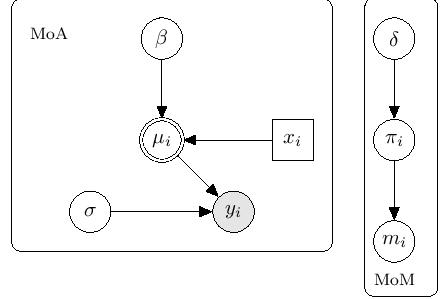}} \hspace{1cm}
\subfloat[Missing At Random (MAR)]{\includegraphics{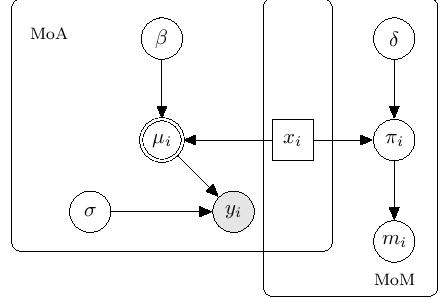}}  \\
\subfloat[Missing Not At Random (MNAR)]{\includegraphics{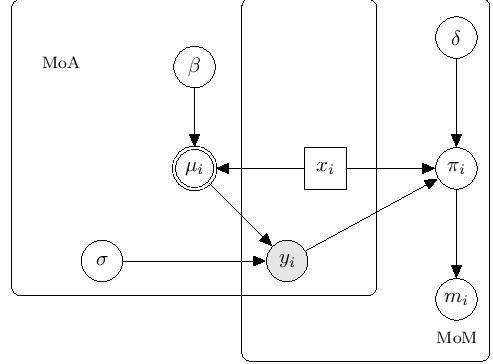}}
\caption{Graphical representation of Rubin's missing data mechanism classes, namely MCAR (a), MAR (b) and MNAR (c). Variables and parameters are represented through nodes of different shapes and colours. Parameters are indicated by grey circles with logical parameters defined by double circles, while predictor variables are assumed fixed and drawn as white squares. Fully observed variables are denoted by white circles, partially observed variables by darker grey circles. Nodes are related to each other through dashed and solid arrows which respectively represent logical functions and stochastic dependence. MoA=Model of Analysis, MoM=Model of Missingness.}\label{FMDM}
\end{figure}

\clearpage

\begin{figure}[!h]
\centering
\subfloat[Missing costs (2003-2009)]{\includegraphics[scale=.45]{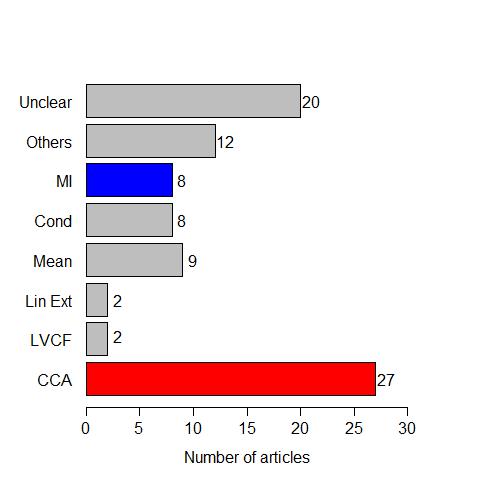}} \hspace{1cm}
\subfloat[Missing costs (2009-2015)]{\includegraphics[scale=.45]{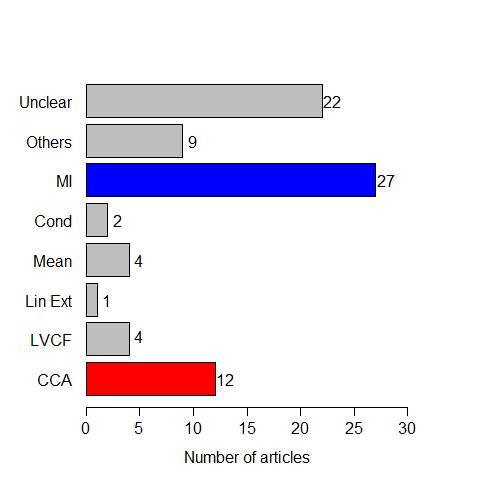}} \\
\subfloat[Missing effects (2003-2009)]{\includegraphics[scale=.45]{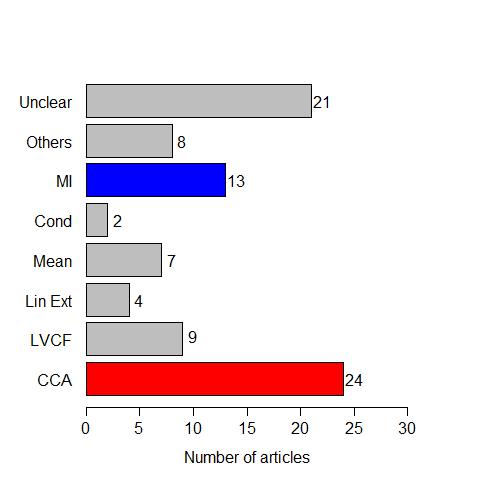}} \hspace{1cm}
\subfloat[Missing effects (2009-2015)]{\includegraphics[scale=.45]{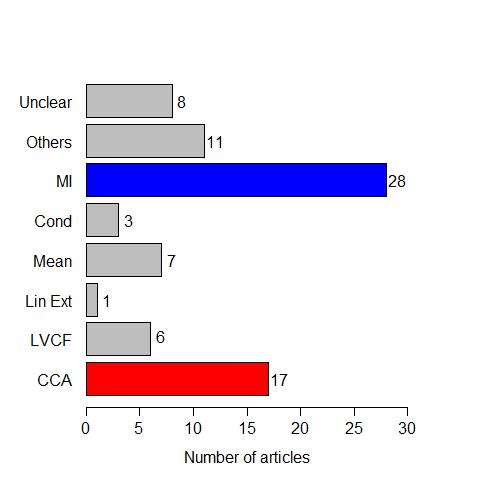}} 
\caption{Review of the base-case methods used to handle missing cost and effect data between 2003-2009 and 2009-2015. Legend: Complete Case Analysis (CCA), Last Value Carried Forward (LVCF), Linear Extrapolation (Lin Ext), Mean Imputation (Mean), Conditional Imputation (Cond), Multiple Imputation (MI), any other method present in less than 4 articles (Others), unspecified method (Unclear). The category Unclear includes those articles for which it was not possible, based on the text, to understand the methodology used to deal with the missingness, while the category Others consists of the following methods: Random draw, Linear Mixed Effects Model, Expectation Maximisation algorithm, Input-Case Analysis, Assumed zeros, Two-part regression. The numbers to the right of the bars in the graphs are the numbers of papers including the corresponding method in the base-case analysis.}\label{F2}
\end{figure}

\clearpage

\begin{figure}[!h]
\centering
\subfloat[Robustness Analyses Costs (2003-2009)]{\includegraphics[scale=.45]{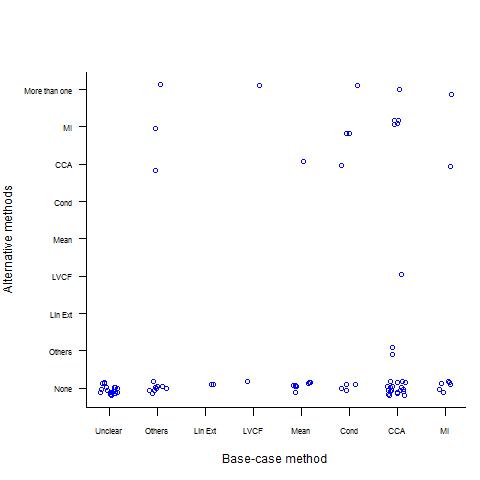}} \hspace{1cm}
\subfloat[Robustness Analyses Costs (2009-2015)]{\includegraphics[scale=.45]{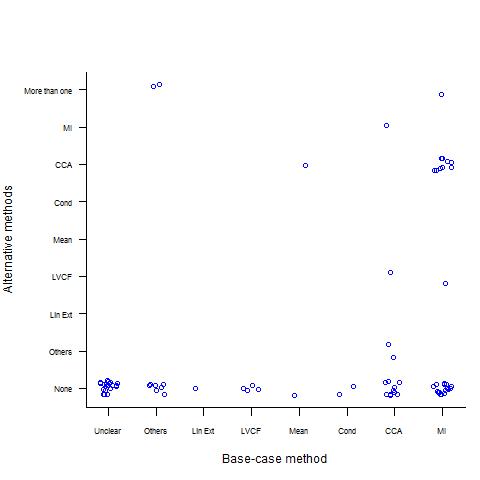}} \\
\subfloat[Robustness Analyses Effects (2003-2009)]{\includegraphics[scale=.45]{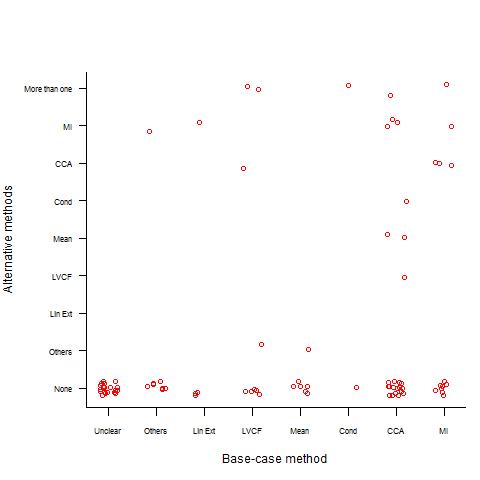}} \hspace{1cm}
\subfloat[Robustness Analyses Effects (2009-2015)]{\includegraphics[scale=.45]{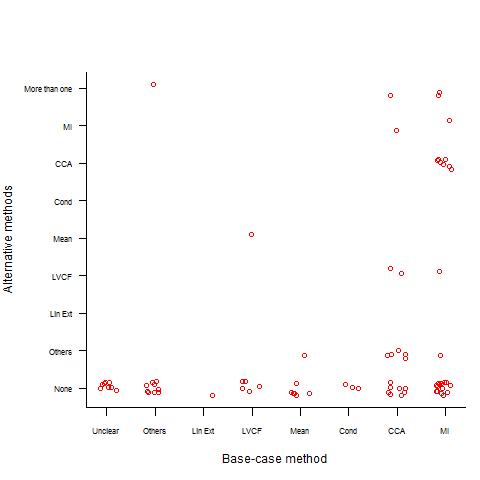}} 
\caption{Comparison of methods used in the base-case analysis ($x$ axis) and those used as alternatives in a robustness analysis ($y$ axis) for the articles between 2003-2009 and 2009-2015 for missing costs and effects. Legend: unspecified methods (Unclear), other methods (Others), Linear Extrapolation (Lin Ext), Last Value Carried Forward (LVCF), Mean Imputation (Mean), Conditional Imputation (Cond), Complete Case Analysis (CCA), Multiple Imputation (MI).The category Unclear includes those articles for which it was not possible, based on the text, to understand the methodology used to deal with the missingness, while the category Others consists of the following methods: Random draw, Linear Mixed Effects Model, Expectation Maximisation algorithm, Input-Case Analysis, Assumed zeros, Two-part regression.}\label{F3}
\end{figure}

\clearpage

\begin{figure}[!h]
\scalebox{0.55}[0.55]{
\includegraphics{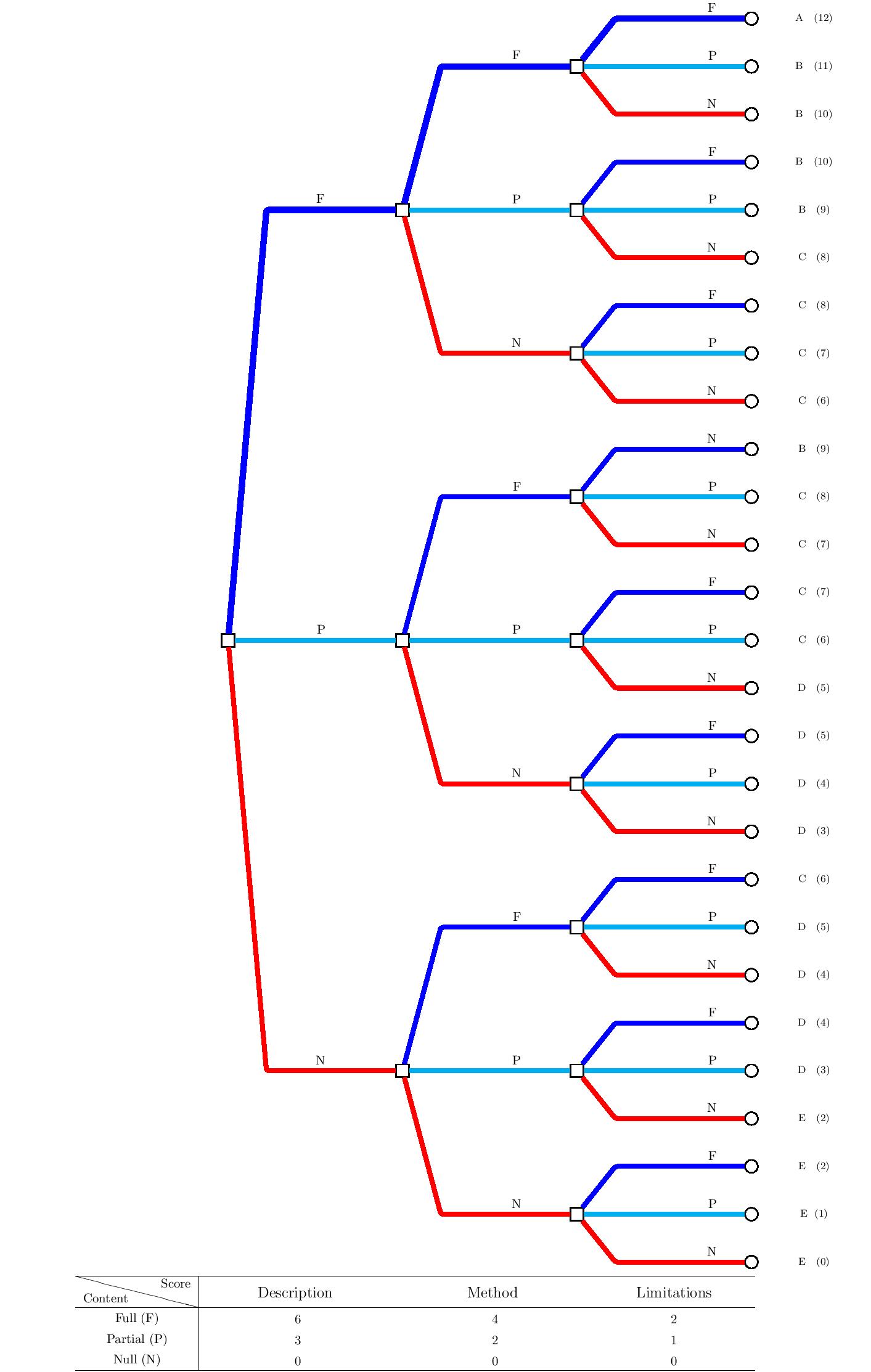}
}\caption{Diagram representation for the quality score categories. The table at the bottom shows how scores have been weighted according to the information provided on each component. In accordance with the table, different edges of the diagram correspond to different components. From left to right, the initial edges are related to the Description, edges in the middle to the Method, and final edges to the Limitations. Edges colour represents the different way the information provided in each analysis component is evaluated: Red=No information (N), Light Blue=Partial information (P), Blue=Full information (F). Final scores (0 - 12) with associated ordered categories (E - A) show the overall level of information provided for each combination of component and content evaluation.
}\label{F6}
\end{figure}

\clearpage

\begin{table}[!h]
\scalebox{0.75}[0.75]{
\noindent
\hskip1.0cm\begin{tabular}[t]{@{}>{\raggedright\arraybackslash}p{0.4\textwidth}}
\center{\textbf{\textit{\large{Description}}}}
\newline
\begin{itemize}[topsep=0pt,itemsep=-2pt,leftmargin=0pt]
\item[] \textbf{Key considerations}
\end{itemize}
\begin{enumerate}[topsep=0pt,itemsep=-2pt,leftmargin=13pt]
\item Report the number of individuals with missing data  for each variable in the reported analysis by treatment group.
\item Describe the missing data patterns for all variables included in the economic analysis (is missingness on one variable associated with missingness on another variable?, is there a longitudinal aspect to the data?)
\item Discuss plausible reasons why values are missing (e.g.~death).
\newline
\end{enumerate}
\begin{itemize}[topsep=0pt,itemsep=-2pt,leftmargin=0pt]
\item[] \textbf{Optimal considerations}
\end{itemize}
\begin{enumerate}[topsep=0pt,itemsep=-2pt,leftmargin=13pt]
\item[1.] Provide supplementary material about the preliminary analysis on missingness  (e.g.~descriptive plots and tables)
\newline\newline
\end{enumerate}
\hskip-2.0cm\begin{tabular}{l}
\hskip2.0cm\textsuperscript{1}\footnotesize{For example, in Multiple Imputation, state the imputation model specification and variables included, the number of imputations, }
\\
\hskip2.0cm\footnotesize{post imputation checks.}
\end{tabular}
\end{tabular}
\begin{tabular}[t]{@{}>{\raggedright\arraybackslash}p{0.4\textwidth}@{}}
\center{\textbf{\textit{\large{Method}}}}
\newline
\begin{itemize}[topsep=0pt,itemsep=-2pt,leftmargin=0pt]
\item[] \textbf{Key considerations}
\end{itemize}
\begin{enumerate}[topsep=0pt,itemsep=-2pt,leftmargin=13pt]
\item Identify a plausible missingness assumption for the specific patterns and setting analysed.
\item State the method and software used in the base-case analysis.
\item For more general methods provide details about their implementation \textsuperscript{1} 
\item Perform a plausible robustness analysis; provide and discuss the results.
\newline\newline\newline\newline
\end{enumerate}
\begin{itemize}[topsep=0pt,itemsep=-2pt,leftmargin=0pt]
\item[] \textbf{Optimal considerations}
\end{itemize}
\begin{enumerate}[topsep=0pt,itemsep=-2pt,leftmargin=13pt]
\item[1.] Provide supplementary material about the method implementation in the base-case and robustness analysis (e.g.~software implementation code)  
\end{enumerate}
\end{tabular}
\begin{tabular}[t]{@{}>{\raggedright\arraybackslash}p{0.4\textwidth}}
\center{\textbf{\textit{\large{Limitations}}}}
\newline
\begin{itemize}[topsep=0pt,itemsep=-2pt,leftmargin=0pt]
\item[] \textbf{Key considerations}
\end{itemize}
\begin{enumerate}[topsep=0pt,itemsep=-2pt,leftmargin=13pt]
\item Acknowledge and quantify the impact of the missing data on the results.
\item State possible weaknesses and issues with respect to the method and assumptions.
\newline\newline\newline\newline
\end{enumerate}
\end{tabular}
}
\caption{List of the information content for each of the three components that we would like to observe in the studies in order to achieve a full analysis reporting of the missing data. The contents are divided into two subgroups: key and optimal considerations. The former are the statements to be considered as mandatory for transparency when conducting an economic evaluation in the presence of missing data. The latter are additional considerations that further extend the analysis reporting of the missing data through supplementary materials. The lack of even one single key considerations is considered to be a partial analysis reporting while a null analysis reporting is related to the absence of all key considerations.}\label{T1}
\end{table}

\clearpage

\begin{figure}[!h]
\centering
\subfloat[Missing Cost Analyses (2009-2015)]{\includegraphics[scale=.45]{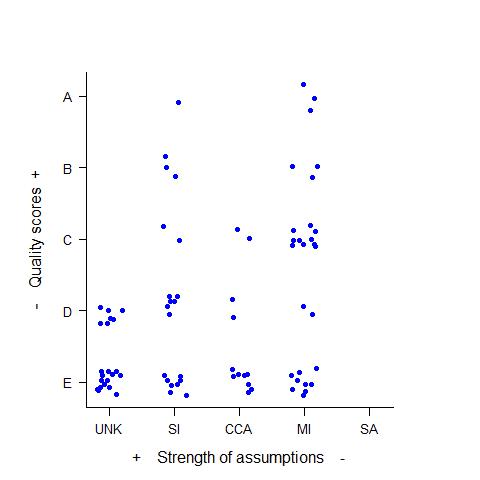}}\label{FAC} 
\subfloat[Missing Effect Analyses (2009-2015)]{\includegraphics[scale=.45]{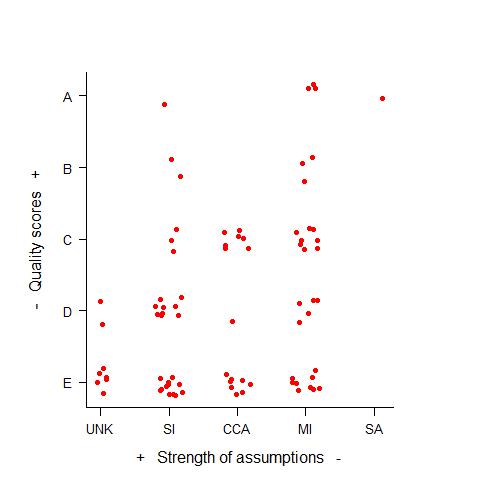}}\label{FAE}
\caption{Joint assessment, in the reviewed articles between 2009-2015, for missing costs and effects, of two components. The x-axis is the missingness method assumptions: Unknown (UNK), Single Imputation (SI), Complete Case Analysis (CCA), Multiple Imputation (MI) and Sensitivity Analysis (SA). The y-axis is the ordered classification for the quality judgement (scores) to support these assumptions: E, D, C, B, A.}\label{F5}
\end{figure}

\clearpage
\bibliographystyle{apa}
\bibliography{Literature_review_paper}

\begin{thebibliography}{}

\bibitem[\protect\astroncite{Baio}{2013}]{Baioa}
Baio, G. (2013).
\newblock {\em Bayesian Methods in Health Economics}.
\newblock Chapman and Hall/CRC, University College London London, UK.

\bibitem[\protect\astroncite{Bang and Robins}{2005}]{Bang}
Bang, H. and Robins, J. (2005).
\newblock Doubly robust estimation in missing data and causal inference models.
\newblock {\em Biometrics}, 61:962--973.

\bibitem[\protect\astroncite{Briggs et~al.}{2003}]{Briggs}
Briggs, A., Clark, T., Wolstenholme, J., and Clarke, P. (2003).
\newblock Missing…. presumed at random: cost-analysis of incomplete data.
\newblock {\em Health Economics}, 12:377--392.

\bibitem[\protect\astroncite{Burton et~al.}{2007}]{Burton}
Burton, A., Billingham, L., and Bryan, S. (2007).
\newblock Cost-effectiveness in clinical trials: using multiple imputation to
  deal with incomplete cost data.
\newblock {\em Clinical Trials}, 4:154--161.

\bibitem[\protect\astroncite{Diaz-Ordaz et~al.}{2014a}]{Diaz-Ordaz2}
Diaz-Ordaz, K., Kenward, M., Cohen, A., Coleman, C., and Eldridge, S. (2014a).
\newblock Are missing data adequately handled in cluster randomised trials? a
  systematic review and guidelines.
\newblock {\em Clinical Trials}, 11:590--600.

\bibitem[\protect\astroncite{Diaz-Ordaz et~al.}{2014b}]{Diaz-Ordaz}
Diaz-Ordaz, K., Kenward, M., and Grieve, R. (2014b).
\newblock Handling missing values in cost effectiveness analyses that use data
  from cluster randomized trials.
\newblock {\em J.R. Statist. Soc.}, 177:457--474.

\bibitem[\protect\astroncite{Faria et~al.}{2014}]{Faria}
Faria, R., Gomes, M., Epstein, D., and White, I. (2014).
\newblock A guide to handling missing data in cost-effectiveness analysis
  conducted within randomised controlled trials.
\newblock {\em PharmacoEconomics}, 32:1157--1170.

\bibitem[\protect\astroncite{Gelman et~al.}{1995}]{Gelman}
Gelman, A., Rubin, D., J., C., and Stern, H. (1995).
\newblock {\em Bayesian data analysis}.
\newblock Chapman and Hall, London, UK.

\bibitem[\protect\astroncite{Graves et~al.}{2002}]{Graves}
Graves, N., Walker, D., Raine, R., Hutchings, A., and Roberts, J. (2002).
\newblock Cost data for individual patients included in clinical studies: no
  amount of statistical analysis can compensate for inadequate costing method.
\newblock {\em Health Economics}, 11:735--739.

\bibitem[\protect\astroncite{Groenwold et~al.}{2012}]{Groenwold}
Groenwold, R., Rogier, A., Donders, T., Roes, K., Harrell, F., and Moons, K.
  (2012).
\newblock Dealing with missing outcome data in randomized trials and
  observational studies.
\newblock {\em American Journal of Epidemiology}, 175:210--217.

\bibitem[\protect\astroncite{Harkanen et~al.}{2013}]{Harkanen}
Harkanen, T., Maljanen, T., Lindfors, O., Virtala, E., and Knekt, P. (2013).
\newblock Confounding and missing data in cost-effectiveness analysis:
  comparing different methods.
\newblock {\em Health Economics Review}, 3.

\bibitem[\protect\astroncite{Lambert et~al.}{2008}]{Lambert}
Lambert, P., Billingham, L., Cooper, N., Sutton, A., and Abrams, K. (2008).
\newblock Estimating the cost-effectiveness of an intervention in a clinical
  trial when partial cost information is available: a bayesian approach.
\newblock {\em Health Economics}, 17:67--81.

\bibitem[\protect\astroncite{Little et~al.}{2010}]{Littlea}
Little, R., D’Agostino, R., Dickersin, K., Emerson, S., Farrar, J.,
  Frangakis, C., Hogan, J., Molenberghs, G., Murphy, S., Neaton, J., Rotnitzky,
  A., Scharfstein, D., Shih, W., Siegel, J., , and Stern, H. (2010).
\newblock The prevention and treatment of missing data in clinical trials.
  panel on handling missing data in clinical trials.
\newblock {\em Committee on National Statistics, Division of Behavioral and
  Social Sciences and Education}.

\bibitem[\protect\astroncite{Manca and Palmer}{2006}]{Manca}
Manca, P. and Palmer, S. (2006).
\newblock Handling missing values in cost effectiveness analyses that use data
  from cluster randomized trials.
\newblock {\em Appl Health Econ Health Policy}, 4:65--75.

\bibitem[\protect\astroncite{McLachlan and Krishnan}{2008}]{McLachlan}
McLachlan, G. and Krishnan, T. (2008).
\newblock {\em The EM Algorithm and Extensions}.
\newblock John Wiley and Sons, Haboken, New Jersey.

\bibitem[\protect\astroncite{Molenberghs et~al.}{2015}]{Molenberghs}
Molenberghs, G., Fitzmaurice, G., Kenward, M., Tsiatis, A., and Verbeke, G.
  (2015).
\newblock {\em Handbook of Missing Data Methodology}.
\newblock Chapman and Hall, New York.

\bibitem[\protect\astroncite{Noble et~al.}{2012}]{Noble}
Noble, S., Hollingworth, W., and Tilling, K. (2012).
\newblock Missing data in trial-based cost-effectiveness analysis: the current
  state of play.
\newblock {\em Health Economics}, 21:187--200.

\bibitem[\protect\astroncite{Oostenbrink and Al}{2005}]{Oostenbrink}
Oostenbrink, J. and Al, M. (2005).
\newblock The analysis of incomplete costdata due to dropout.
\newblock {\em Health Economics}, 14:763--776.

\bibitem[\protect\astroncite{Powney et~al.}{2014}]{Powney}
Powney, M., Williamson, P., Kirkham, J., and Kolarnunnage-Dona, R. (2014).
\newblock Multiple imputation to deal with missing eq-5d-3l data: Should we
  impute individual domains or the actual index?
\newblock {\em Trials}, 15.

\bibitem[\protect\astroncite{Ramsey et~al.}{2005}]{Ramsey}
Ramsey, S., Willke, R., Briggs, A., Brown, R., Buxton, M., and Chawla, A.
  (2005).
\newblock Good research practices for cost-effectiveness analysis alongside
  clinical trials: the ispor rct-cea task force report.
\newblock {\em Value Health}, 8:521--533.

\bibitem[\protect\astroncite{Richardson and Manca}{2004}]{Richardson}
Richardson, G. and Manca, A. (2004).
\newblock Calculation of quality adjusted life years in the published
  literature: a review of methodology and transparency.
\newblock {\em Health Economics}, 13:1203--1210.

\bibitem[\protect\astroncite{Rombach et~al.}{2016}]{Rombach}
Rombach, I., Rivero-Arias, O., Gray, A., Jenkinson, C., and Burke, O. (2016).
\newblock The current practice of handling and reporting missing outcome data
  in eight widely used proms in rct publications: a review of the current
  literature.
\newblock {\em Qual Life Res}.

\bibitem[\protect\astroncite{Rubin}{1987}]{Rubina}
Rubin, D. (1987).
\newblock {\em Multiple Imputation for Nonresponse in Surveys}.
\newblock John Wiley and Sons, New York,USA.

\bibitem[\protect\astroncite{Schafer}{1997}]{Schafera}
Schafer, J. (1997).
\newblock {\em Analysis of Incomplete Multivariate Data}.
\newblock Chapman and Hall, New York,USA.

\bibitem[\protect\astroncite{Schafer}{1999}]{SchaferMI}
Schafer, J. (1999).
\newblock Multiple imputation: a primer.
\newblock {\em Statistical Methods in Medical Research}, 8:3--15.

\bibitem[\protect\astroncite{Schafer and Graham}{2002}]{SchaferMD}
Schafer, J. and Graham, J. (2002).
\newblock Missing data: Our view of the state of the art.
\newblock {\em Psychological Methods}, 7:147--177.

\bibitem[\protect\astroncite{Simons et~al.}{2015}]{Simons}
Simons, C., Arias, O., Yu, L., and Simon, J. (2015).
\newblock Multiple imputation to deal with missing eq-5d-3l data: Should we
  impute individual domains or the actual index?
\newblock {\em Qual Life Res}, 24:805--815.

\bibitem[\protect\astroncite{White and Carlin}{2010}]{White}
White, I. and Carlin, J. (2010).
\newblock Bias and efficiency of multiple imputation compared with
  complete-case analysis for missing covariate values.
\newblock {\em Statistics in Medicine}, 29:2920--2931.

\bibitem[\protect\astroncite{Wood et~al.}{2004}]{Wood}
Wood, A., White, I., and Thompson, S. (2004).
\newblock Are missing outcome data adequately handled?a review of published
  randomized controlled trials in major medical journals.
\newblock {\em Clinical Trials}, 1:368--376.

\end{thebibliography}

\appendix
\renewcommand\bibsection{\section*{\refname{: Review 2003-2009}}}
\includepdf[pages={1-}]{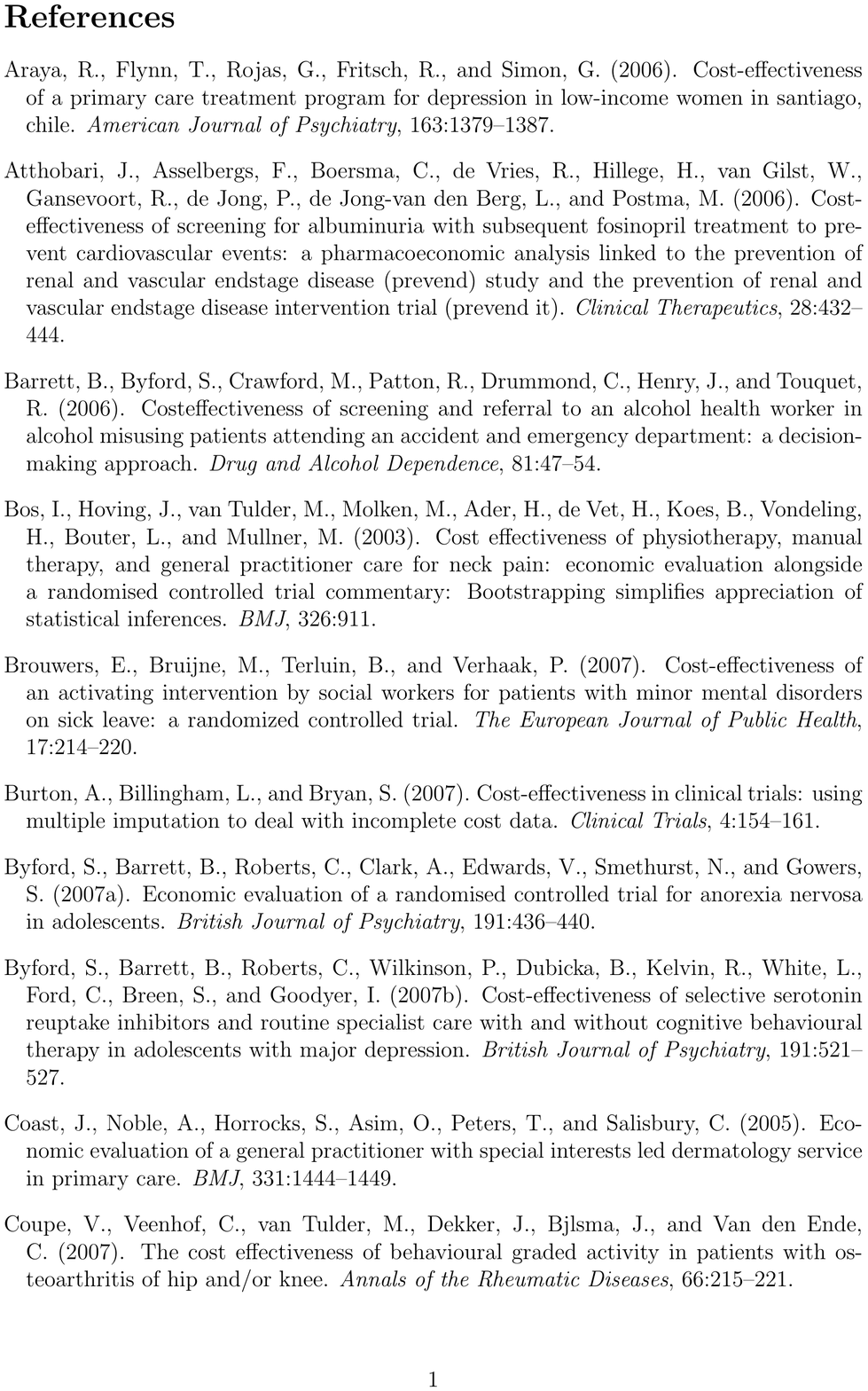}
\renewcommand\bibsection{\section*{\refname{: Review 2009-2015}}}
\includepdf[pages={1-}]{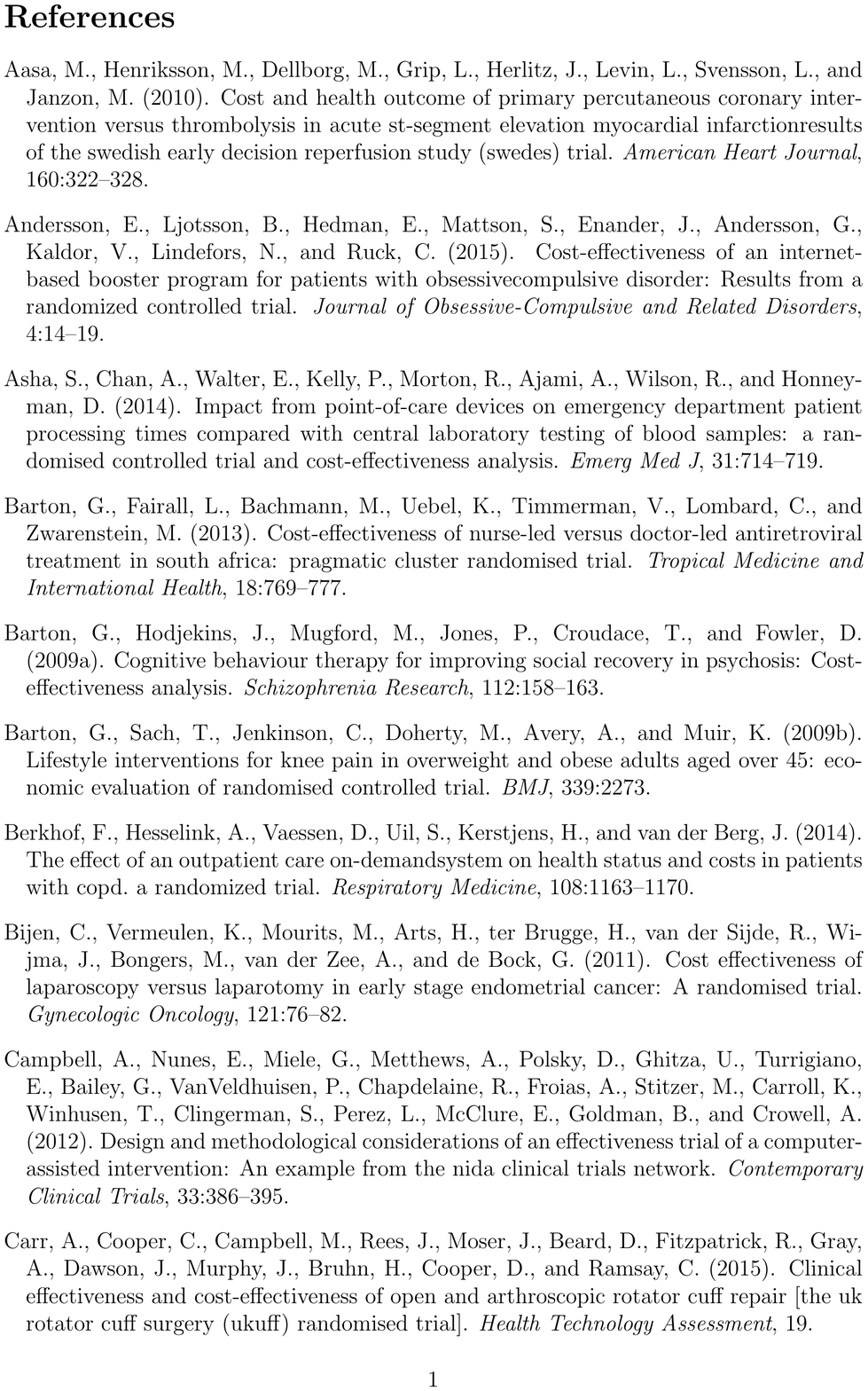}

\end{document}